\documentstyle[12pt,a4wide]{article}
\begin{document}

\begin{center}
{\Large \bf Dynamical Evolution of Galaxy Groups.\\ A comparison of two 
approaches.}

\vspace*{0.5cm}
{\large Garc\'{\i}a-G\'omez, C.\footnote[1]{Universitat Rovira i Virgili. 
E.T.S.E. Carretera Salou s/n. 43006 Tarragona Spain}, Athanassoula, E. 
\footnote[2]{Observatoire de Marseille, 13248 Marseille Cedex 4, France} 
and Garijo, A.\footnotemark[1]} \end{center}

{\bf Abstract.}
In this paper we test the performance of explicit simulations of groups 
of galaxies, (i.e simulations in which each galaxy is treated as a mass 
point and the physics of the interactions is modelled by specific 
analytical prescriptions for merging conditions), by comparing them with 
fully self-consistent simulations starting from identical initial 
conditions. The quality of the explicit simulations is very unequal. For 
some prescriptions the results are in complete disagreement with the 
self-consistent simulations. The inclusion of other dynamical effects 
like dynamical friction gives, in some cases, better agreement. We also 
propose a new merging criterion, which, combined with dynamical friction, 
gives much better agreement with self-consistent simulations in a variety 
of initial conditions, but even this criterion has a limited range of 
applicability. 

\section{Introduction.}

Fully self-consistent N-body simulations, where each galaxy is 
represented by a large number of particles, are a useful, albeit 
expensive,
tool for studying the evolution of galaxy groups and clusters. However, 
for simulations of large clusters of galaxies, like the Coma cluster, the 
necessary computing time is prohibitive. As a substitute people have 
consi\-dered explicit simulations, in which each galaxy is represented by a 
single point and the physics of the interactions is modelled by explicit 
prescriptions for merging conditions. In particular,
a variety of recipes are explored for the conditions the two galaxies 
must fulfill in order to merge. In general, these merging conditions are 
based on self-consistent si\-mulations of two-galaxy collisions, and do not 
include the tidal
forces between the galaxies or collisions involving more than two 
galaxies. It is thus not a priori certain that they will perform well in 
simulations of group or cluster evolution. In some cases (Merritt, 1983; 
Richstone and Malumuth, 1983; Mamon 1987), the authors also introduce 
other effects like dynamical friction and tidal forces from the
background. The main advantage of this type of approach is that it is 
inexpensive in computing time and therefore allows one to explore a wide
parameter space. In any case, a considerable fraction of the results on 
the dynamics of galaxy groups are
based on the explicit approach. We may cite works by Jones and Efstathiou 
(1979), Roos and Norman (1979), Aarseth and Fall (1980), Cooper and 
Miller (1981), Roos (1981), Roos and Aarseth (1982), Merritt (1983), 
Richstone and Malumuth (1983), Malumuth and Richstone (1984), Saarinen 
and Valtonen (1985), Mamon (1987), Navarro et al. (1987) and Schindler 
and B\"ohringer (1993).

Not many self-consistent simulations of groups with more than 10 galaxies 
can be found in the literature. We can cite the articles by Carnevalli et 
al. (1981), Ishizawa et al. (1983), Ishizawa (1986), Rhee and Roos 
(1990), Barnes (1992), Funato et al. (1993) and Bode et al. (1994). The 
first works of this kind used Aarseth's (1971) N-body code and a limited 
number of points, typically $10-20$, to
represent each galaxy, and only recently it has become possible to use 
the order of 1000 particles per galaxy.

Our aim is to compare the two approaches to see whether, and under what 
conditions, one can use explicit simulations and have confidence in the 
results. For this purpose, we have evolved a set of initial conditions in 
two different ways. One way is to use an N-body code where physics is 
included explicitly, the other, to use self-consistent simulations and a 
treecode (Barnes and Hut 1986, Hernquist 1987 for a vectorised 
version), representing each galaxy either by $100$ or by $900$ points. 
In section 2 we describe our initial conditions and the different merging 
criteria used so far in the literature. In section 3 we compare the 
results of fully self-consistent numerical simulations to those of 
explicit simulations made with the various merging criteria, both without 
(section 3.1) and with dynamical friction (section 3.2). This comparison 
led us to propose a new merging criterion (section 3.3), whose 
performance we also compare with the fully self-consistent simulations. 
In this section we consider only groups with no common all-encompassing
dark matter halo. Simulations including such a halo are presented in 
section 4, where again we compare the results of self-consistent and 
explicit simulations. We summarise and discuss our results in section 5.

\section{Initial conditions and merging criteria} 

We have considered five different initial conditions, labeled A, B, C, D 
and H, each for systems consisting of 50 galaxies. In simulations A, B, D 
and H the radial distances from the
galaxy centers to the center of the group were picked at random between 0 
and $R_{out}$.
For simulation C the central part of the sphere contained no galaxy, i.e. 
the radial distances were picked between $0.5R_{out}$ and $R_{out}$. For 
simulations
A to D all the mass is in the individual galaxies, while in simulation H 
we included a common live halo, centered on the center of the group, and 
containing half of the total mass. The halo density distribution is a 
Plummer one with a core radius equal to half $R_{out}$. Run A starts in 
free-fall, and we will often refer to it as the collapsing group. The 
velocity dispersions in the remaining three runs were chosen to be 
independent of radius, gaussian, isotropic, and such that the system of 
galaxies starts off in virial equilibrium. Simulation D is similar 
to B but more compact, as the radius of the sphere containing all the 
galaxies is half that of run B. The particles in a given galaxy were 
initially taken to follow a Plummer distribution of core radius equal to 
0.2 and of unit mass. When evolved in isolation, an individual
galaxy first shows a low amplitude relaxation in the very first few
time steps due to the fact that the simulations have a softening, while 
the analytical Plummer sphere does not. After that, and for a time equal 
to that during which the group simulations were run, the galaxies do not
evolve any further. Thus, during that time, for the representations with 
900 particles per
galaxy, the radii containing 25\%, 50\% and 75\% of the mass of the 
galaxy vary only by a couple of
percent. For the representation with 100 particles the radii containing 
25\% and 50\% of the mass vary by 4-5\%, and only the radius containing 
75\% of the mass varies significantly, particularly in the later phases 
of the evolution.

More
information on the initial conditions for the simulations is summarised 
in Table~1.
{\sl Column}~1 contains the name of the simulation, {\sl Column}~2 gives 
$R_{out}$, the
radius of the sphere containing the group at the start of the simulation, 
{\sl Column}~3 shows the initial mean separation between the galaxies and 
{\sl Column}~4 gives the ratio between the initial velocity dispersion of 
the galaxies considered as point masses, $\sigma_{cl}$, divided by the 
velocity dispersion of the particles within a single galaxy, 
$\sigma_{gal}$. {\sl Column}~5 contains the crossing time defined as
\begin{equation}
t_{cr}=\left( {\frac{2R_h^3}{GM}}\right) ^{1/2}, \end{equation}

\noindent where $R_h$ represents the half mass radius. Finally, {\sl 
Column}~6 contains the ratio between $t_{tot},\/$ the total duration of 
the simulation and $t_{cr}$. All through this paper our units are such 
that the gravitational constant $G~=~1$. 

\begin{table}
\begin{center}
\caption{Initial conditions of the simulations} \vskip 0.25cm
\begin{tabular}{llllll}
\hline
Run & $R_{out}$ & $<sep>$ & $\sigma _{cl}/\sigma _{gal}$ & $t_{cr}$ & 
$t_{tot}/t_{cr}$ \\ \hline
A & 30 & 8.0 & 0.0 & 15.3 & 2.0 \\
B & 20 & 6.8 & 1.4 & 4.5 & 6.7 \\
C & 20 & 10.3 & 1.0 & 11.6 & 2.6 \\
D & 10 & 3.4 & 1.9 & 1.6 & 18.7 \\
H & 20 & 7.4 & 2.7 & 8.9 & 3.4\\\hline
\end{tabular}
\end{center}
\end{table}

The self-consistent simulations were run using the vectorised version 
(Hernquist 1988) of the Barnes-Hut tree algorithm (Barnes and Hut 1986), 
with a softening of 0.05 and an opening angle $\theta=0.7$. In explicit 
simulations each galaxy is represented by a single point to which is 
associated a mass, an internal energy and a core radius. These parameters 
may change during the evolution of the system due to the different 
interactions suffered by the point-galaxies and we used the recipes of 
Aarseth and Fall (1980) to follow their time evolution. The explicit 
simulations are of course much faster than the self-consistent
ones. A complete self-consistent simulation with $900$ points per galaxy 
took $521663$ seconds in a Cray YMP 2L computer. The self-consistent 
simulation with $100$ particles per galaxy lasted only the $5\%$ of this 
time and the explicit simulation only $0.2\%$. 

In order to compare the results of the different kind of simulations we 
consider the time evolution of the following global parameters of the 
groups:

\begin{enumerate}
\item Number of galaxies: $N_{gal}$
\item Half mass radius: $R_h$, where $M(R_h)=1/2\,M_{tot}$ \item Three 
dimensional velocity dispersion: \end{enumerate}

$$\sigma_v^2=\sum_{i=1}^{N_{gal}}\frac{m_i\mid {\bf v_i}-<{\bf v}> 
\mid^2}{M_{tot}-m_i(t=0)},\,\,
{\rm where}\,\,<{\bf v}>=\sum_{i=1}^{N_{gal}}\frac{m_i {\bf v_i}} 
{M_{tot}}.$$

\noindent where all quantities are evaluated at each timestep, except for 
$m_i(t=0) = 1$ which is the mass of all individual galaxies at the 
start of the simulations and is taken to be $m_i(t=0) = 1$.

In our explicit simulations we consider, in a first stage, only merging 
between galaxies. In a second set of simulations we include also the 
effect of dynamical friction. In this way we can check the importance of 
both effects. Merging between galaxies is usually described in the 
literature using an explicit condition involving the separation and 
relative velocities of the pair of galaxies. If this condition is 
fulfilled, the two galaxies are merged in a single one in this timestep, 
taking into account the conservation of mass, energy and momentum 
(Aarseth \& Fall 1980). If this condition is not fulfilled both galaxies 
survive and continue their motion. 

We found in the literature various criteria which have been used to 
decide whether two galaxies are going to merge and we used all of them in 
turn in our explicit simulations. The condition of Roos and Norman (1979, 
hereafter condition RN) is: \begin{equation}
v(r_p)\leq 3.1\sigma (1-0.3\frac{r_p}{R_g}) 
\left(\frac{1+m_2/m_1}{2}\right)^{1/4}
\end{equation}

\noindent
where $m_2 \leq m_1$ and $r_p/R_g < 1$. $r_p$ is the minimum separation 
between the galaxies, $v(r_p)$
is their relative velocity at $r_p$, and $R_g$ is the larger of their 
radii. This criterion was obtained empirically from collisions between 
galaxies described by fewer than $100$ particles. 

Aarseth and Fall (1980, hereafter condition AF) used the criterion: 
\begin{equation}
{\left[ \frac{r_p}{{2.6(\epsilon _1+\epsilon _2)}}\right] }^2+ 
{\left[\frac{v(r_p)}{{1.16v_e(r_p)}}\right] }^2\leq 1,
 \end{equation}

\noindent
which is a simple fit to the results of the simulations of van Albada and 
van Gorkom (1977), White (1978) and Roos and Norman (1979). The core 
radius of galaxy {\it i} is $\epsilon _i$, while $v_e(r_p)$ is the escape 
velocity of the system composed of the two galaxies before merging at 
pericenter:
\begin{equation}
v_e^2(r_p) = 2 G (m_1 + m_2)(r_p^2 + \epsilon_1^2 + \epsilon_2^2)^{-1/2}.
\end{equation}

Farouki and Shapiro (1982, hereafter condition FS) obtained a similar 
condition for the merging of two rotating galaxies with massive halos and 
spins aligned with the orbital angular momentum: \begin{equation}
{\left[ \frac{r_p}{{5.5(\epsilon _1+\epsilon _2)}}\right] }^2+ {\left[ 
\frac{v(r_p)}{{1.1v_e(r_p)}}\right] }^2\leq 1. \end{equation}

This condition predicts more mergings than the criterion from Aarseth and 
Fall (1980) for two reasons. It favours collisions between galaxies 
further apart and it forces the spins to be aligned. This criterion is 
not directly applicable to our case, where we use initially nonrotating 
Plummer spheres, but we include it for the sake of completeness.

Finally Richstone and Malumuth (1983, hereafter condition RM) use the 
different criterion:
\begin{equation}
r_pv(r_p)\leq \left[ 8/3\,G^2\,(m_1<r_2^2>\,+\,m_2<r_1^2>)(m_1+m_2) 
\right] ^{1/4},
\end{equation}

\noindent
which is a generalisation of a criterion propo\-sed by Tre\-maine (1980)
for the case of different masses. The value $<r_i^2>$ is the mean 
quadratic radius of a galaxy. For the case of a Plummer sphere 
$<r^2>=\epsilon ^2/2$, and this is the value we have used in our 
simulations.

To save computer time we do not need to apply the adopted merging 
criterion to all galaxy pairs at all times. Following Navarro {\it et 
al.} (1987), we check whether the condition is fulfilled only if the 
separation between two galaxies is smaller than $3(r_{h_1}+r_{h_2})$, 
where $r_{h_i}$ is the half mass radius of the galaxy {\it i}. This 
separation is sufficiently large so that merging events are not missed, 
while speeding up considerably the computations. 

As the simulations evolve a central giant ``galaxy" is formed as a result 
of the mer\-gings and/or tidal stripping of the galaxies in the group. 
Dynamical friction between this and the remaining individual galaxies 
influences the
evolution and we have therefore included this effect in the explicit 
simulations, using the well known
Chandrashekar (1943) formula for the deceleration: \begin{equation}
{\bf a_v} = - \frac{4\pi G^2\, m_{gal} \ln \Lambda \rho({\bf r})} {v^3} 
F(v){\bf v}
\end{equation}
where
\begin{equation}
F(v) = erf(X) - \frac{2X}{\sqrt{\pi}}e^{-X^2} \end{equation}
and $erf(X)$ is the error function, $X=v/\sqrt{2}\sigma$, $\sigma$ is the 
velocity dispersion of the
objects in the background, and $m_{gal}$ is the mass of the galaxy 
travelling at speed ${\bf v}$;
$\rho({\bf r})$ is the density of the central galaxy, considered as a 
Plummer sphere, at the position of the
secondary galaxy, ${\bf r}$ being the relative separation of their 
centers, and $\Lambda = b_{max} / b_{min}$, where $b_{max}$ and $b_{min}$ 
are the maximum and minimum
impact parameters of encounters contributing to the drag. When we include 
a common halo we apply Eq. (6) twice, once for the central giant 
``galaxy" and the other for the halo, adding these two accelerations. 

The self-consistent simulations where analyzed as follows. First, we 
need to define the central giant ``galaxy", which we will refer to in 
this paragraph simply as the central object. In order to do so, we 
analyze at each timestep
separately each subsystem composed of the particles that were bound at 
$t=0$ in a single galaxy. Using the positions and velocities of these 
particles we discard from the subsystem all particles with positive 
energy
relative to it and consider them as part of the central object. The 
particles that still form a bound subsystem will define the state of the 
galaxy at this timestep. If after this process a galaxy contains less 
than $10\%$ of the particles it had at $t=0$, we discard this subsystem 
as a galaxy and we add all its particles to the central object, thus 
considering that the initial galaxy has been definitely disrupted. For 
each of the remaining galaxies we use the $35\%$ of its most bound 
particles to define its position and velocity. Finally, we also consider 
possible mergings between the remaining galaxies, as well as between 
these galaxies
and the central object. Two galaxies were merged in a single one if the 
following conditions are satisfied:
\begin{eqnarray*}
\Delta r & < & a(r_{c1}+r_{c2}) \\
\Delta v & < & b(\sigma_1 + \sigma_2)
\end{eqnarray*}
\noindent where $r_{ci}$ is the radius of the sphere containing the 
$35\%$ most bound particles and $\sigma_i$ its velocity dispersion. The 
constants $a=1.4$
and $b=0.6$ were selected in order to have smooth central objects. The 
parameters of this object were calculated with just the $10\%$ most bound 
particles and not with the $35\%$ as with the rest of the galaxies. This 
ensured that we do not consider a merger between the central galaxy and 
another galaxy while they still form two separate objects. We finally 
used the positions and velocities of the remaining galaxies
to define the global parameters of the system.

\section{Simulations without common dark matter halo} 

\subsection{Evolution without dynamical friction.} 

In Fig.~1 we show the evolution of the number of galaxies $N_{gal}$ as a 
function of time for all the simulations without distributed dark matter. 
In the first column we compare the self-consistent simulations with $900$ 
and $100$ particles per galaxy with the explicit simulations obtained 
with the AF and RM conditions.
In the second column, the evolution of the number of galaxies in the 
self-consistent simulations is compared with the explicit simulations 
using the FS and RN conditions. 

We note that the explicit simulations perform rather unequally. The 
results depend on the type of initial conditions and on the mergng 
condition used to describe the interactions. Globally we can say that the 
AF and RM conditions
seem to follow the time evolution of $N_{gal}$ much better than the FS 
and RN conditions. In the first stages of the evolution of the collapsing 
group (Run A), the less tightly bound and virialised group (Run B) and 
for Run C, which is a virialised group with no central mass 
concentration, both AF and RM conditions describe the time evolution of 
the number of galaxies rather well. This is not true, however, for Run D 
(tightly bound and virialised group), for which the AF condition 
overestimates the number of mergings from the start, while the RM 
condition does the opposite. As the evolution proceeds the discrepancies 
between the self-consistent simulations and the explicit simulations 
become more evident. For all initial conditions the FS and RN conditions 
overestimate the number of mergings from the start. The sole exception is 
the explicit simulation with the FS condition in the case of Run C, where 
the agreement with the self-consistent simulation is quite good. 

For the time evolution of the half-mass radius, $R_{h}$, we find similar 
results. This can be seen in Fig.~2, where the panels refer to the same 
initial conditions as in Fig.~1. In general, the explicit simulations 
controlled by the AF and RM conditions show a better global behaviour 
than the simulations governed by the FS and RN conditions. This is due to 
the high number of mergings predicted by the latter conditions. In the 
case of Run A, all explicit simulations follow well the collapse phase. 
When most of the mass is accumulated in the central area, the number of 
encounters is relatively large and there are strong interactions with the 
giant central galaxy. At this moment, self-consistent and explicit 
simulations separate. The AF and RM conditions allow some galaxies to 
avoid merging with the giant galaxy in the first passage and the system 
experiences an expansion which is not shown in self-consistent 
simulations. On the other hand, the FS and RN conditions predict a much 
higher rate of mergers than the self-consistent simulations and we are 
left too early with only a single giant galaxy. In the case of Run~B, the 
AF and RM conditions describe very well the state of the system during 
the first part of the simulations. As the simulation evolves, however, 
some galaxies reach the central parts where they suffer an hyperbolic 
encounter with the central mass concentration of the giant galaxy instead 
of merging with it, as is the case in the self-consistent simulations, 
because the merging criteria strongly disfavour merging in high speed 
collisions. This makes the system expand, an effect which is not seen in 
the self-consistent simulations. This does not happen for Run C where 
there is no such central mass concentration and explicit and 
self-consistent simulations follow the same evolution, except for minor 
differences and a strong deviation for the case of condition RN. Run D is 
the most difficult case for the explicit simulations. In this situation 
galaxies move at higher speeds than in Run B or Run C. Surprisingly, in 
the case of self-consistent simulations, this does not make merging with 
the central object more difficult, as one might expect naively in the 
first instance. However, the RM conditions predicts more hyperbolic 
encounters than the self-consistent simulations, giving strong 
oscillations of the half mass radius. On the other hand, the AF condition 
seem to describe the situation quite well. The number of galaxies 
predicted by the RN and FS conditions
are well below the numbers predicted by the self-consistent simulations, 
again due to the high number of mergers predicted by these conditions.

Finally, in Fig.~3 we show similar comparisons, now for the three 
dimensional velocity dispersion.
The larger number of mergers predicted by the FS and RN conditions nearly 
always gives lower velocity dispersions than the self-consistent 
simulations as well as strong oscillations due to small number
statistics. On the other hand, the AF and RM conditions give a better 
general description of the evolution of the three dimensional velocity 
dispersion. This is specially true for Run A, where all the
motion is nearly radial and only small discrepancies appear at the end of 
the simulations. For the case of Run B and the RM condition, the 
hyperbolic encounters which lead to a higher half mass radius of the 
system, give also higher velocity dispersions, because some galaxies 
which merge in the self-consistent simulations can escape in the explicit 
ones. The AF condition describes this time evolution much better. The 
velocity dispersion of Run C is well described for both conditions until 
shortly before the end of the simulation, when both conditions predict 
higher velocity dispersions than the self-consistent simulations. In the 
case of Run D the RM condition has again
some difficulty in describing the behaviour of the self-consistent 
simulations. This is also due to the high number of large deflections of 
the secondary galaxies. The AF condition follows well the evolution of 
the three dimensional velocity dispersion in this situation. 

We would like to note at this point that the self-consistent simulations 
with $100$ particles per galaxy and with $900$ particles per galaxy do 
not show major differences. The number of galaxies as a function of time 
does not change appreciably between these two simulations and this for 
all the initial conditions, i.e. both for virialised and collapsing 
groups. In this sense our results differ from those of van Kampen (1995), 
who found that the small virialised clumps formed during the simulations associated with the 
galaxies do not resist the passage through the central part of the 
cluster. This could be due to the somewhat lower number of particles per 
galaxy, since the typical galaxies in van Kampen's simulations are 
composed of 10-50 points (van Kampen 1995).

Similarly good agreement between the 900 and 100 points per galaxy 
simulations is found for the
velocity dispersion. Somewhat bigger differences, in particular for run 
B, can be seen for the half-mass radius, but even these are not 
excessive.

\subsection{Simulations with dynamical friction.} 

Figure~4 compares the time evolution of the number of galaxies in the 
self-consistent simulations and in the explicit simulations when the 
effect of dynamical friction is included. Since this slows down the 
galaxies and thus
favours merging, the number
of galaxies, $N_g$ will diminish faster. This is clearly seen in all the 
panels of Fig.~4. As can be seen from the left hand panels, this worsens 
the predictions of the RN and FS conditions. The right hand panels show 
that the agreement is now better for the RM condition, and worst for the 
AF one. For the case of Run A there is a systematic deviation between the 
AF condition and the self-consistent simulations. On the other hand, the 
RM condition which had, in the absence of dynamical friction, predicted a 
low number of mergings is, in this case, in much better agreement with 
the self-consistent case. The same can be said about Run B, while in Run 
C the effect of dynamical frictions is not noticeable. This is not 
surprising as we take into account only the effect of dynamical friction 
with the most massive galaxy which, in this case, is practically 
nonexistent. For the most difficult case, Run D, the AF condition falls 
below the results of the self-consistent simulations while the RM 
condition gives good agreement. 

The evolution of the half mass radius is also affected by the inclusion 
of dynamical friction, as is shown in Fig.~5, where we plot the evolution 
of $R_h$ as a function of time. For the explicit simulations with the RN 
and FS conditions dynamical friction does not alter the strong 
disagreement with the
self-consistent simulations. This happens because the explicit 
simulations with these conditions allow too many mergings and we are left 
with a single supergiant galaxy at the center of the system which 
contains a large fraction of the mass and some small satellites. On the 
other hand, there is now a much better agreement between the explicit 
simulations made with the AF and RM conditions and the self-consistent 
cases. For Run A neither condition shows a secondary bouncing of the 
system. The dynamical
friction acts as a braking mechanism that favours merging between the 
secondary galaxies and the central one and a lower number of satellites 
survive in this situation. In Run B, the hyperbolic encounters of the 
satellite galaxies with the central giant are not present and there is no 
later expansion of the system as in the explicit simulations without 
dynamical friction. The explicit simulations with both
the AF and RM conditions predict too small a half mass radius. For Run C, 
as there
is no giant galaxy, dynamical friction is unimportant and all the 
simulations again show the same general behaviour. In Run D the galaxies 
move faster because the system is more tightly bound. The explicit 
simulations with the RM condition and no dynamical friction were not 
capable of describing the evolution of the self-consistent simulations. 
The inclusion of dynamical friction gives a much better agreement between 
these two simulations. On the other hand, the explicit simulations with 
the AF conditions seem to be systematically below the predictions of the 
self-consistent simulations. 

As can be seen in Fig.~6, the three dimensional velocity dispersion shows 
marked diffe\-rences between the explicit simulations and the 
self-consistent ones. As was the case in the absence of dynamical 
friction the explicit simulations with the RN and FS conditions do not 
track well the self-consistent results. The dynamical friction effect is 
barely noticeable in this case, except for some tendency towards lower 
velocity dispersions. As the RN and FS conditions predict many mergings, 
we are left with a giant galaxy in the center and a low number
of satellites orbiting around it. The dispersions are then low but they 
are more subject to fluctuations and have stronger oscillations. 
Including dynamical friction in the explicit simulations with the AF and 
RM conditions does not substantially improve their
results as can be seen if we compare Fig.~6 with Fig.~3. For runs A and C 
the situation is further improved and the explicit simulations follow the 
self-consistent ones very well. Bigger differences between the explicit 
simulations with and without dynamical friction are found for the 
virialised groups (Run B and D). The values predicted by the AF condition 
are now always near the values obtained with the self-consistent 
simulations. However, this is not the case for the RM parametrization. 
For Run B, there are marked differences between these explicit 
simulations and the last phase of the self-consistent simulations. For 
the case of Run D the RM condition gives a systematically higher velocity 
dispersion than the self-consistent simulations.

\subsection{A new merging criterion}

As we have seen, none of the merging criteria proposed so far in the 
literature is capable of describing the time evolution of the global 
properties of groups of galaxies in the variety of situations considered 
in this paper. We can
say that, in general, the AF and RM conditions perform better that the FS 
and RN ones, but even they
fail to describe the evolution of some of the groups. This has motivated 
our search for a more adequate merging criterion. 

We searched for a formula of a form similar to the one proposed by 
Aarseth and Fall (1980), namely:
\begin{equation}
{\left[ \frac{(m_1+m_2)r_p}{{a(m_1\epsilon _1+m_2\epsilon _2)}} 
\right]}^2+{\left[\frac{v(r_p)}{{bv_e(r_p)}}\right] }^2\leq 1. 
\end{equation}
For the part concerning the velocities, we keep the same expression as in 
the Aarseth and Fall formula, which performs quite well in the case of 
the time evolution of the three dimensional velocity dispersions. For the 
part concerning the cores of the galaxies and the separation at 
pericenter we use a mass weighted expression with the aim of taking into 
account possible differences in collisions between galaxies of different 
masses as in the expression due to Richstone and Malumuth (1983). The 
constants $a$ and $b$ are free parameters and will be determined using 
the self-consistent simulations as a reference. This expression can be 
viewed as the equation of the points within an ellipse centered at the 
origin in the plane defined by $(m_1+m_2)r_p/ (m_1\epsilon_1+ m_2\epsilon 
_2)$ and $v(r_p)/v_e(r_p)$. Then $a$ and $b$ are the semimajor axes of 
this ellipse. Increasing the value of $a$ means increasing the axis of 
the ellipse corresponding to the relative separation at pericenter and 
thus allowing mergings in more distant collisions. On the other hand, 
if we increase the value of $b$ we allow merging in faster 
collisions. With this in mind, we fitted the values of $a$ and $b$ to 
the self-consistent simulations using as the basis for our exploration 
the values used by Aarseth and Fall (1980). After some trials and 
comparisons with the self-consistent simulations we obtained the 
following merging criterion: \begin{equation}
{\left[ \frac{(m_1+m_2)r_p}{{2.5(m_1\epsilon _1+m_2\epsilon _2)}} \right] 
}^2+
{\left[\frac{v(r_p)}{{1.18v_e(r_p)}}\right] }^2\leq 1. \end{equation}
The effect of this new criterion is shown in Figs.~7, Fig.~8 and Fig.~9, 
where we compare the time evolution of the global parameters of the 
self-consistent simulations with that of the explicit simulations using 
the AF and RM criteria and our new one. The dynamical friction with the 
most massive galaxy is also included in these cases. 

In Fig.~7 we show the time evolution of the number of galaxies $N_g$. In 
the first column, we repeat the comparison between the self-consistent 
simulations
and the explicit simulations with the AF and RM criteria and dynamical 
friction. In the second column, we have the comparison between the 
self-consistent simulations and the explicit simulations with dynamical 
friction and our new merging criterion. As can be seen, while the 
explicit simulations with the RM criterion mimic quite well the 
self-consistent simulations, this is not true for the AF condition. On 
the other hand, our new criteria follows quite well the evolution of the 
number of galaxies given by the self-consistent simulations for all 
initial conditions.

In Fig.~8 we show the time evolution of the half mass radius. For the 
case of Run A both AF and RM conditions follow quite well the 
self-consistent simulations until the point of maximum collapse. After 
this point, the half mass radius given by these explicit simulations 
falls below the
self-consistent case. Our new condition, however, follows the 
self-consistent simulations with $900$ particles very well. For the case 
of Run B, the AF and RM conditions end below the self-consistent case. 
Our new criterion performs better, following the self-consistent 
simulations, but with some oscillations. For runs C and D we can say that 
all three criteria give similar results. 

Figure~9 which gives the time evolution of the three dimensional velocity

dispersion, is the most interesting one. We have seen that the AF and RM 
conditions give good results for the case of the collapsing group (Run A) 
and this is true also for our new criterion. However, the AF and RM 
explicit simulations do not work well for the case of a virialised group 
(Run B). The AF condition ends with a higher velocity dispersion and the 
RM with a smaller velocity disperson compared to the self-consistent 
case; on the other hand, our new criterion performs much better than 
either. This is specially true for the most difficult case, Run D, the 
virialised and tightly bound group. In this case our new criterion 
performs much better than the AF and RM criteria.

\section{Simulations with a dark matter halo encompassing the whole 
group}

Several observations suggest that clusters and groups of galaxies 
may contain much matter not bound to the galaxies. This led us to run a
self-consistent simulation (Run H), where part of the mass of the system 
is distributed in a background. In the corresponding explicit
simulations the background is included as a rigid Plummer potential with 
the same parameters as the live background in the initial conditions of 
the self-consistent
simulation. The explicit simulations include dynamical friction with the 
most massive galaxy and with the Plummer halo. 

The evolution of the group leads to a system where the central part of 
the galaxy distribution has contracted, while the outer one has expanded. 
This results in an increase of the half-mass radius and a lowering of the 
velocity dispersion, as shown in Fig.~10. The upper panels give the time 
evolution of the number of galaxies in the system $N_g$, the middle ones 
that of the half mass radius $R_h$ and the lower ones that of the three 
dimensional velocity dispersion. In the left panels the self-consistent 
simulations are compared to
the explicit simulations with the AF and RM conditions and in the right 
panels with simulations using our new criterion. As we can see, the 
number of
galaxies diminishes slower in simulations including a common halo than in 
the case of virialised
simulations with no distributed dark matter. The AF and RM conditions 
underestimate the real number of mergers, and so, though to a lesser 
extent, does our new criterion.
For the time evolution of the half mass radius there are strong 
discrepancies between the self-consistent simulations and the explicit 
simulations using any of the merging criteria including the new 
criterion proposed in the previous section. 

The three dimensional velocity dispersion of the galaxies is well 
described by the explicit simulations using any of the merging criteria. 
This global parameter systematically decreases during the simulation as 
the galaxies that move faster near the center disappear and form the 
giant central object. The slope of this evolution flattens off toward the 
end of the simulations. This behaviour is not well followed by the 
explicit simulations using the AF or RM criterion. On the other hand, 
Fig.~10 shows that our new merging criterion is able to reproduce these 
minor details better.

\section{Summary.}

In this paper we compared self-consistent simulations of galaxy groups 
with simulations where the physics of the interactions is modelled by 
merger rules. We used two sets of self-consistent simulations, one in 
which the galaxies were modelled with 900 points and the other with 100 
points. Insofar as the global dynamical
parameters are concerned, the evolution of galaxy groups is similar in 
those two cases. This shows that simulations with a relatively low number 
of particles can be used to follow the evolution of global dynamical 
properties of groups or clusters. However, from the work of van Kampen 
(1995)
it can be inferred that using lower that 100 points per galaxy can be 
dangerous.

As far as the explicit simulations are concerned, we show that the 
conditions used in the literature to
simulate the merging between galaxies are of unequal quality. Of these 
conditions, in the case
where there is neither dynamical friction nor tidal forces, the best are 
those of Tremaine (1980), modified for the case of different masses by 
Richstone and Malumuth (1983), and the one by Aarseth and Fall (1980). 
When we include dynamical friction effects the AF condition predicts too 
many mergers but still maintains good predictions for the rest of the 
global parameters. The condition proposed by Richstone and Malumuth 
(1980) does better as far as the number of galaxies and $R_h$ are 
concerned, but considerably worse for the velocity dispersion. 

As none of these criteria seems to be a good guide for the time evolution 
of the groups as compared with the self-consistent simulations, we have 
fitted a new
criterion to the results of self-consistent simulations. This new 
criterion is:
\begin{equation}
{\left[ \frac{(m_1+m_2)r_p}{{2.5(m_1\epsilon _1+m_2\epsilon _2)}} \right] 
}^2+
{\left[\frac{v(r_p)}{{1.18v_e(r_p)}}\right] }^2\leq 1,
 \end{equation}
and is inspired in the expressions given by Aarseth and Fall (1980) and 
Richstone and Malumuth (1980). This new criterion mimics relatively well 
the time evolution of the global parameters of the groups in as wide a
variety of situations as those presented by our simulations A to D. 
However it performed not so well in case H which has a common halo, but 
this can be explained by the different nature of the simulations  
implying that even this new criterion has only a limited range of 
applicability.

Our comparisons show that some of the older results on the dynamics 
of groups and clusters of galaxies should be viewed with caution. For 
instance, Roos (1981) studied the evolution of expanding systems of 
galaxies to simulate the evolving universe. As he used the RN criterion 
in his simulation the predicted merger rate can be too high. In the same 
way, when Roos and Aarseth (1982) used this criterion to study the 
evolution of the luminosity function of a cluster of galaxies, their 
final luminosity functions can be artificialy peaked towards
high luminositues. Similarly, Valtonen et al. (1984), Saarinen and 
Valtonen (1985) and Perea et al. (1990) use explicit simulations to 
criticize the virial mass obtained for galaxy clusters. We have, however, 
seen that this kind of simulation is biased toward higher velocity 
dispersions. Finally, the explicit simulations on compact groups by Mamon 
(1987) using a diffuse intergalactic background may also be biased.

Thus we can conclude that there is no ideal substitute for fully 
self-consistent N-body simulations. However, in cases when one needs to 
look only at global quantities describing the system and is not 
interested in fine structure and details, a first exploration of 
parameter space can be done using explicit simulations and the criterion 
proposed in this paper. This performs particularly well in cases where 
the group has no common halo.

{\bf Acknowledgements.}
We thank Albert Bosma and Kevin Prendergast for reading and improving the 
manuscript, and our referee, Joshua Barnes for his useful 
suggestions and criticism which improved the quality of this paper. We 
also thank L. Hernquist for making available to us his vectorised version 
of the treecode.
Some of the simulations discussed in this paper were made at the C98 of 
the IDRIS (Institut du d\'eveloppement et des ressources en informatique 
scientifique, Orsay, France). 

\noindent
{\Large \bf References.}

\noindent
Aarseth, S.J. 1971 ApSS 14,20\\
Aarseth, S.J., Fall, S.M. 1980 ApJ 236,43\\ Barnes, J. 1992 in {\sl 
Morphological and physical classification of galaxies} G. Longo et 
al.\linebreak \hspace*{0.5cm} (eds.) Kluwer Academic Publishers, 
p277-292\\ Barnes, J., Hut, P. 1986 {\sl Nature} 324,446\\ Bode, P.W., 
Berrington, R.C., Cohn, H.N., Lugger, Ph. M. 1994 ApJ 433,479\\
Carnevalli, P., Cavaliere, A., Santangelo, P. 1981 ApJ 249,449\\ 
Chandrasekhar, S. 1943 ApJ 97,255\\
Cooper, R.G., Miller, R.H. 1982 ApJ 254,16\\ Farouki, S.M., Shaphiro, 
S.L. 1982 ApJ 259,103\\ Funato, Y., Makino, J., Ebisuzaki, T. 1993 PASJ 
45,289\\ Hernquist, L. 1987 ApJS 64,715\\
Ishizawa, T. 1986 ApSS 119,221\\
Ishizawa, T., Matsumoto, R., Tajima, T., Kageyama, H., Sakai, H. 1983 
PASJ 35,61\\
Jones, B.J.T., Efstathiou, G. 1979 MNRAS 189,27\\ Malumuth, E.M., 
Richstone, D.O. 1984 ApJ 276,413\\ Mamon, G.A. 1987 ApJ 321,622\\
Merritt, D. 1983, ApJ 264,24\\
Navarro, J.F., Mosconi, M.B., Lambas, D.G. 1987 MNRAS 228,501\\ Perea, 
J., del Olmo, A., Moles, M. 1990 A\&A 237,328\\ Rhee, G., Roos, N. 1990 
MNRAS 243, 629\\ Richstone, D.O., Malumuth, E.M. 1983 ApJ 268,30\\ Roos, 
N. 1981 A\&A 95,349\\
Roos, N., Aarseth, S.J. 1982 A\&A 114,41\\ Roos, N., Norman, C.A. 1979 
A\&A 76,75\\ Saarinen, S., Valtonen, M.J. 1985 A\&A 153,130\\ Schindler, 
S., B\"ohringer, H. 1993 A\&A 269,83\\ Tremaine, S.D. 1980 {\sl The 
Structure and Evolution of Normal Galaxies.} ed. S.M. Fall et D. 
\linebreak \hspace*{0.5cm} Lynden-Bell. Cambridge University Press\\ van 
Albada, T.S., van Gorkom, J.H. 1977 A\&A 54,121\\ van Kampen, E. 1995 
MNRAS 273,295\\
Valtonen, M.J., Innanen, K.A., Huang, T.-Y., Saarinen, S. 1985 A\&A 
143,182\\ White, S.D.M. 1978 MNRAS 184,185\\

\newpage

\noindent
{\Large \bf Figure Captions.}

\noindent {\bf Fig.~1} Comparison of the time evolution of the number of 
galaxies in the self-consistent simulations with $100$ particles per 
galaxy (thin line) and $900$ particles per galaxy (thick line) with the 
explicit simulations for the same initial conditions and without 
dynamical friction. In the left panels we use the AF and RM merging 
conditions and in the right panels we use the FS and RN ones. The initial 
conditions of each simulation are described in Table~1.

\noindent {\bf Fig.~2} Comparison of the time evolution of the half mass 
radius of the system in self-consistent simulations and in explicit 
simulations without dynamical friction for the same initial conditions. 
The symbols are as in Fig.~1. 

\noindent {\bf Fig.~3} Time evolution of the three dimensional velocity 
dispersion of the galaxies consi\-dered as point masses for the 
self-consistent and the explicit simulations. The symbols are as in 
Fig.~1.

\noindent {\bf Fig.~4} Time evolution of the number of galaxies $(N_g)$ 
in the self-consistent simulations compared with the evolution of this 
number in the explicit simulations with dynamical friction included. The 
thick lines correspond to the self-consistent simulations with $900$ 
particles per galaxy and the thin lines to the simulations with $100$ 
points per galaxy. In the first column, we show the comparison with the 
explicit simulations using the AF criterion and using the RM criterion. 
In the second column, we show the same comparisons with the explicit 
simulations using the FS condition and using the RN condition. 

\noindent {\bf Fig.~5} Same as for Fig.~4 but for the time evolution of 
the half mass radius of the system.

\noindent {\bf Fig.~6} Same as for Fig.~5 but for the time evolution of 
the three dimensional velocity dispersion. 

\noindent {\bf Fig.~7} Comparison of the explicit simulations using the 
AF and the RM criteria with the explicit simulations using the new 
criterion. The performance of each criterion is compared with the 
self-consistent simulations. In the first column, we show the time 
evolution of the number of galaxies in the self-consistent simulations 
with $900$ particles per galaxy (thick lines) and with $100$ particles 
per galaxy (thin lines) compared with the explicit
simulations using the AF criterion and using the RM criterion. In the 
second column, we compare the time evolution of $N_g$ for the 
self-consistent simulations with the results of the explicit simulations 
using the new criterion. In all cases we include dynamical friction.

\noindent {\bf Fig.~8} Same as Fig.~7 but for the time evolution of the 
half mass radius of the system.

\noindent {\bf Fig.~9} Same as Fig.~7 but for the time evolution of the 
three dimensional velocity dispersion. 

\noindent {\bf Fig.~10} Time evolution of the global parameters of the 
simulations with distributed background. In both columns we show the 
evolution of $N_g$, $R_h$ and $\sigma (3D)$ for the self-consistent 
simulations with $450$ particles per galaxy (thick lines) and for the 
self-consistent simulations with $100$ particles per galaxy (thin lines). 
In the left panel these are compared with the explicit simulations with 
the AF condition and with the RM condition. In the right panel the 
self-consistent simulations are compared with the explicit simulations 
with our new criterion. 

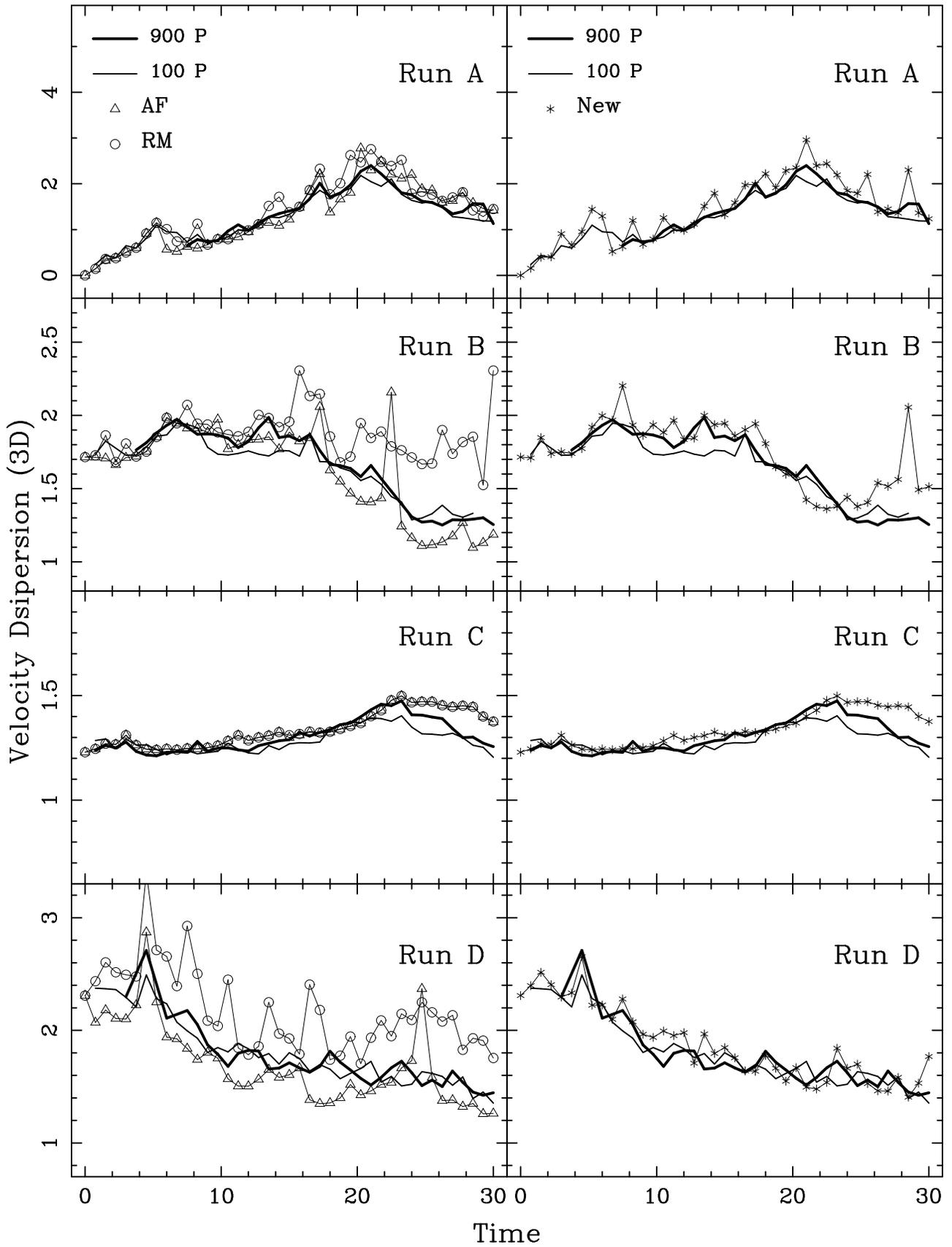
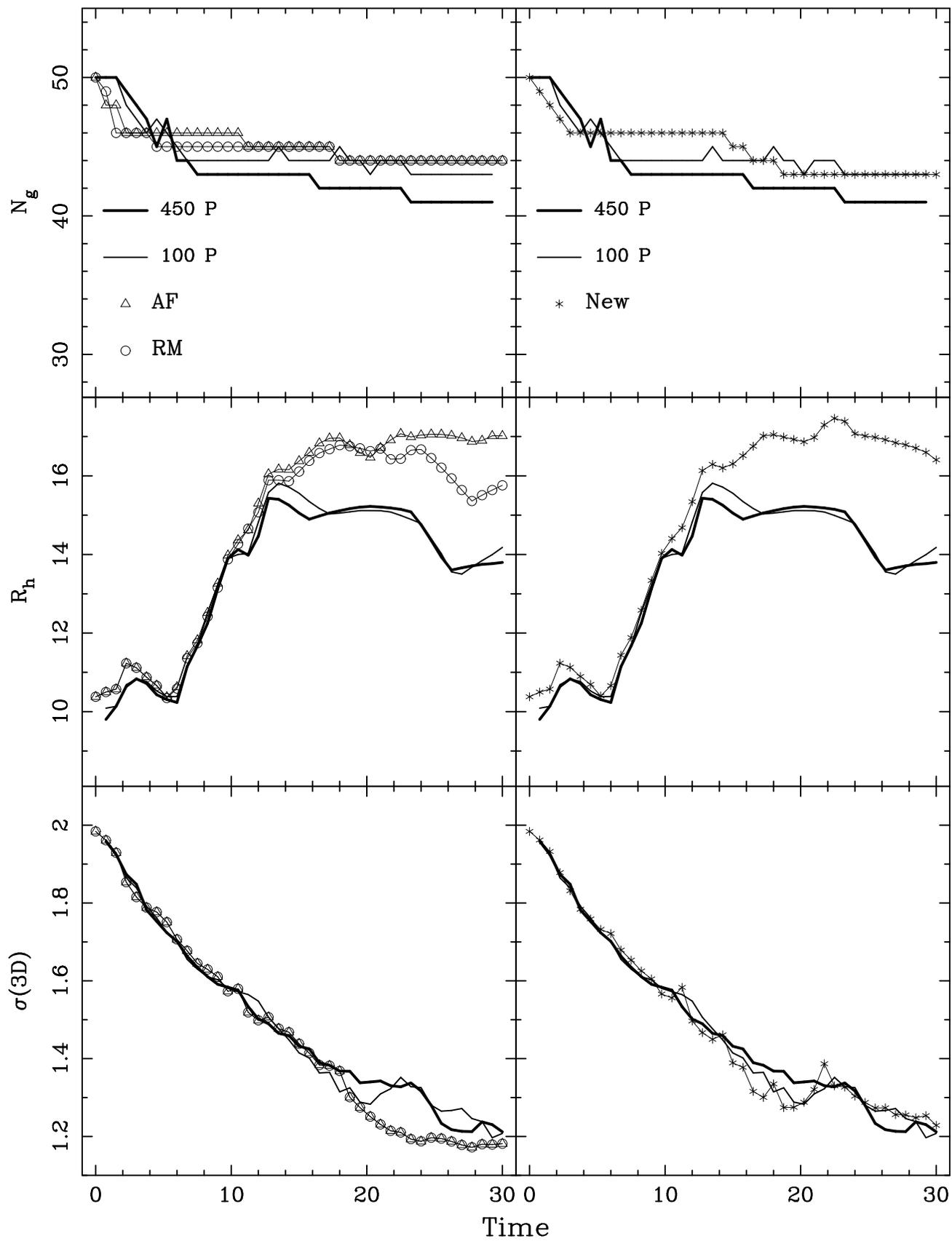
\end{document}